\documentclass{article}

\usepackage[final]{neurips_2023}
\usepackage{hyperref}
\usepackage{bbding}

\usepackage{natbib}
\setcitestyle{numbers,square}

\usepackage[utf8]{inputenc} 
\usepackage[T1]{fontenc}    
\usepackage{hyperref}       
\usepackage{url}            
\usepackage{booktabs}       
\usepackage{amsfonts}       
\usepackage{nicefrac}       
\usepackage{microtype}      
\usepackage{xcolor}         
\usepackage{multirow}
\usepackage{graphicx}
\usepackage{CJKutf8}
\usepackage[many]{tcolorbox}
\usepackage{graphicx}

\usepackage{geometry}
\usepackage{tabularx}
\usepackage{booktabs}
\usepackage{makecell}

\title{OpenDataLab: Empowering General Artificial Intelligence with Open Datasets}

\author{Conghui He, Wei Li, Zhenjiang Jin, Chao Xu, Bin Wang, Dahua Lin\thanks{Corresponding author} \\
Shanghai Artificial Intelligence Laboratory\\
Shanghai, 200232, China\\
    \texttt{\{heconghui,liwei,jinzhenjiang,xuchao,wangbin,lindahua\}@pjlab.org.cn} \\}
\begin{document}

\maketitle

\begin{abstract}

The advancement of artificial intelligence (AI) hinges on the quality and accessibility of data, yet the current fragmentation and variability of data sources hinder efficient data utilization. The dispersion of data sources and diversity of data formats often lead to inefficiencies in data retrieval and processing, significantly impeding the progress of AI research and applications. To address these challenges, this paper introduces OpenDataLab, a platform designed to bridge the gap between diverse data sources and the need for unified data processing. OpenDataLab integrates a wide range of open-source AI datasets and enhances data acquisition efficiency through intelligent querying and high-speed downloading services. The platform employs a next-generation AI Data Set Description Language (DSDL), which standardizes the representation of multimodal and multi-format data, improving interoperability and reusability. Additionally, OpenDataLab optimizes data processing through tools that complement DSDL. By integrating data with unified data descriptions and smart data toolchains, OpenDataLab can improve data preparation efficiency by 30\%. We anticipate that OpenDataLab will significantly boost artificial general intelligence (AGI) research and facilitate advancements in related AI fields. For more detailed information, please visit the platform's official website: \href{https://opendatalab.com/}{https://opendatalab.com/}.

  
\end{abstract}

\section{Introduction}

In the rapidly evolving field of artificial intelligence, data has become an indispensable resource~\cite{deng2009imagenet,lin2014microsoft,schuhmann2022laion,changpinyo2021conceptual}. However, the isolation between datasets and the incompatibility with data processing tools have been major technical barriers hindering the widespread application and advancement of AI technologies. To address these challenges, OpenDataLab has adopted an innovative technological architecture aimed at eliminating barriers between datasets and enhancing data accessibility and operability, thereby facilitating comprehensive advancements in AI.

\begin{figure*}[h]
    \centering
    \noindent\includegraphics[width=1.0\textwidth]{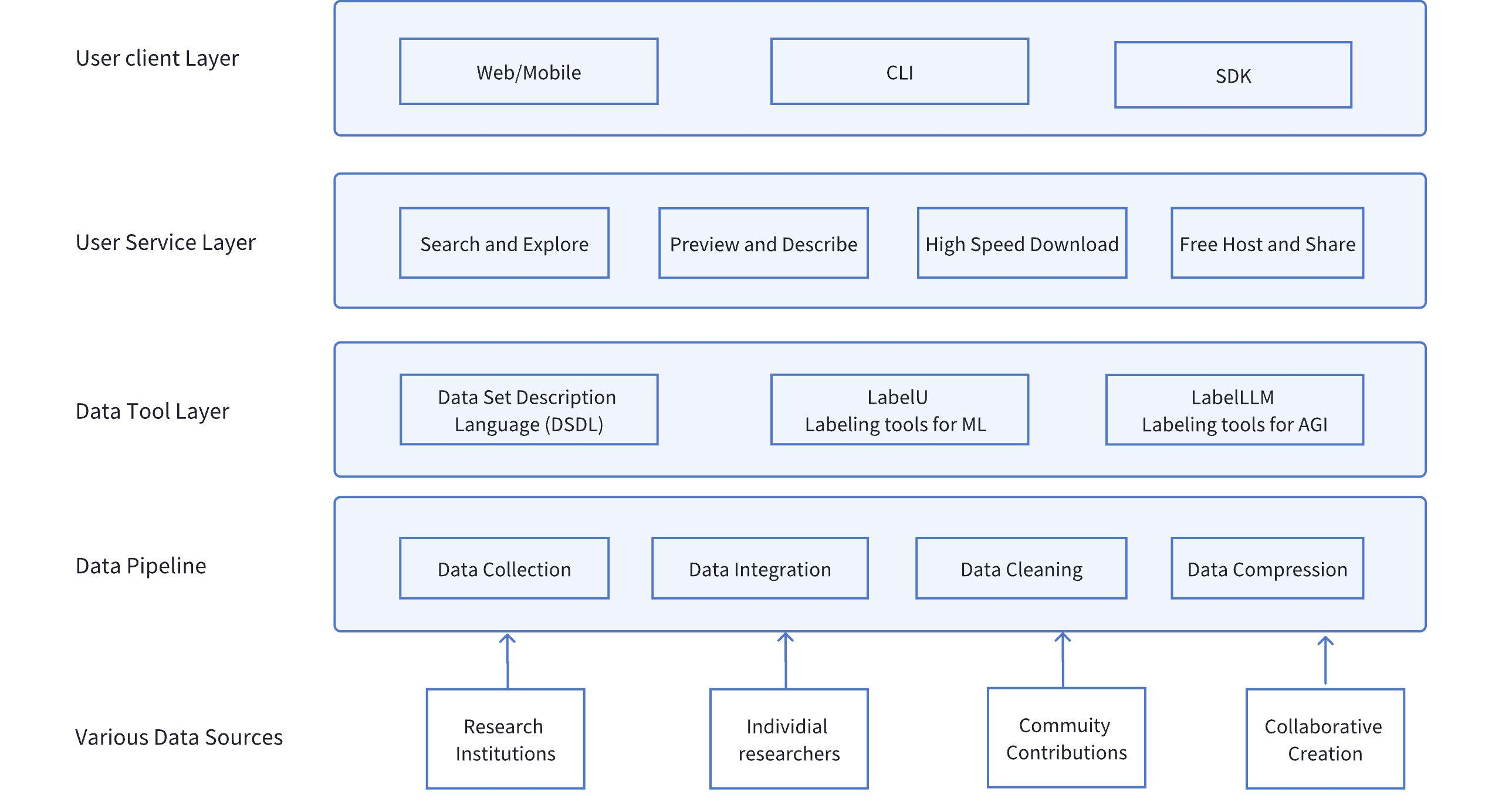}
    \caption{Architectural Framework of OpenDataLab Platform}
    \label{fig:figure1}
\end{figure*}

The architectural design of OpenDataLab is meticulously constructed, as illustrated in Figure~\ref{fig:figure1}, and includes the following layers from bottom to top:
\begin{itemize}
    \item \textbf{Diverse Data Sources}: Encompassing data from research institutions, individual researchers, community shares, and collaborative constructions, ensuring comprehensiveness and diversity of data.

    \item \textbf{Data Pipeline Layer}: Incorporates critical steps such as data collection, integration, cleaning, and compression, ensuring high quality and consistency of data.

    \item \textbf{Data Tools Layer}: Features the proprietary Data Set Description Language (DSDL~\cite{wang2024dsdl}) and annotation tools based on machine learning and AGI technologies, LabelU and LabelLLM, significantly enhancing the efficiency and precision of data processing.

    \item \textbf{User Services Layer}: Offers efficient query retrieval, data visualization, high-speed downloads, and free hosting and sharing services, greatly enhancing user experience.

    \item \textbf{User Client Layer}: Includes a Graphical User Interface (GUI), Command Line Interface (CLI), and Software Development Kit (SDK), catering to different user habits and technical needs.

\end{itemize}

Through this multi-layered, structured architecture, OpenDataLab not only effectively resolves the issue of data silos but also establishes an integrated platform that supports the entire process from data acquisition to application. This integrated solution is highly practical for data scientists and AI researchers, and also provides robust technical support for developers and industry experts.

The introduction of DSDL not only enhances the interoperability between datasets but also simplifies the integration process with data tools, effectively bridging the gap between datasets and tools for seamless data interfacing and efficient utilization. Furthermore, by integrating advanced data annotation tools, OpenDataLab further strengthens its capabilities in data processing and analysis, providing users with a more powerful and flexible data handling environment.

In summary, through the standardization of DSDL and the integration of advanced data tools, OpenDataLab not only addresses the longstanding issue of data silos but also builds bridges between previously incompatible datasets and tools. This innovative technological framework signifies that OpenDataLab is poised to lead a new paradigm in data acquisition and processing in the AI field, driving continuous development and innovation across the industry.

\section{Open Datasets}
\subsection{High-Quality, Diverse Datasets}

\begin{figure*}[h]
    \centering
    \noindent\includegraphics[width=1.0\textwidth]{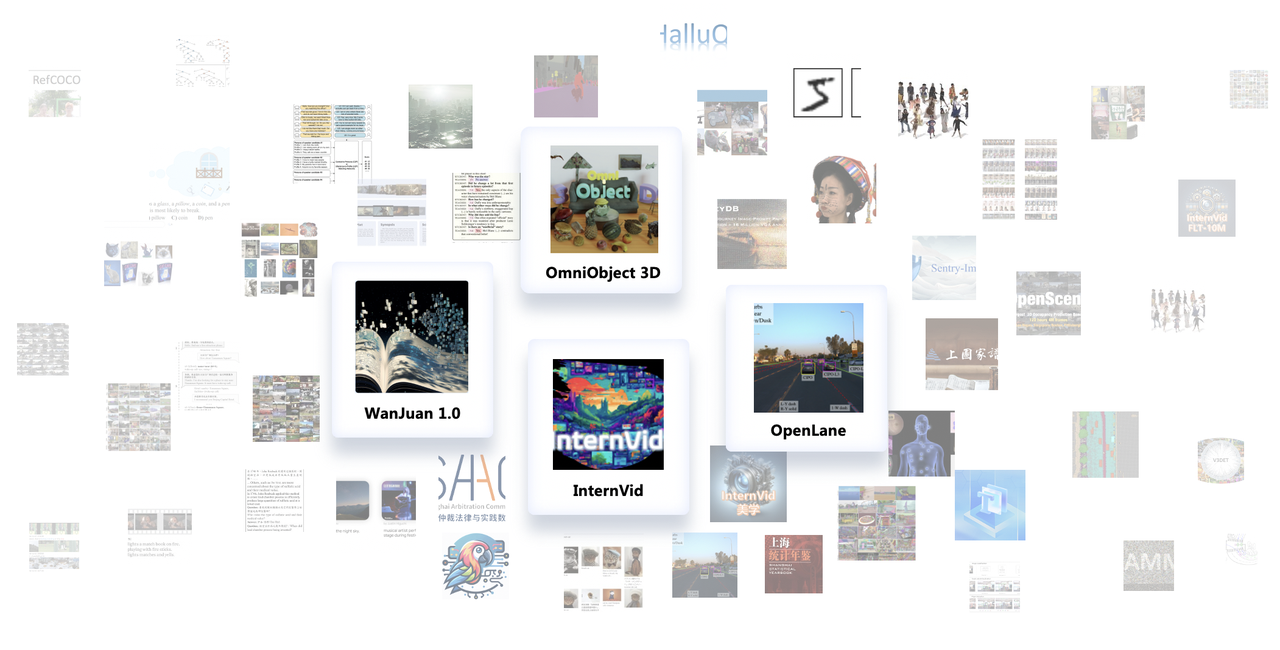}
    \caption{Visualization of Popular Datasets Available at OpenDataLab}
    \label{fig:figure2}
\end{figure*}

OpenDataLab boasts over 6,500 open datasets, covering more than 30 data formats and supporting over 50 types of tasks, offering an extensive range of data topics. These datasets include over 6 billion images, 800 million video clips, 1 trillion tokens, 1 million 3D models, and 20,000 hours of audio, totaling over 80TB of data. This provides users with large-scale, high-quality, and diverse training samples. Visualizations of some popular datasets are shown in Figure~\ref{fig:figure2}.

The platform's data sources are diverse, encompassing open-source data from various academic publications, as well as collections from public dataset platforms and contributions from individual users. OpenDataLab provides comprehensive dataset support for all stages of large model development, from pre-training and fine-tuning to evaluation. These include large-scale pre-training datasets like WanJuan~\cite{he2023wanjuan}, image-text pair dataset Laion5B~\cite{schuhmann2022laion}, and video-centric multimodal dataset InternVid~\cite{wang2023internvid}.

\subsection{Licensing Protection for Datasets}
OpenDataLab's management of datasets includes not only the collection and organization of data but also the systematic organization and verification of dataset metadata. This metadata includes the original sources, official websites, relevant scholarly articles, publishers, and applicable licenses. Through this approach, OpenDataLab offers a free and convenient public data hosting service, ensuring transparency and traceability of data.

Moreover, the platform supports a variety of open licenses to facilitate the broad use and sharing of data. These licenses include Creative Commons (CC), Open Data Commons (ODC), and Community Data License Agreement (CDLA). By clearly outlining the permissions, restrictions, and conditions of each license, OpenDataLab helps users understand and adhere to the specific legal and ethical guidelines required when using datasets, such as attribution (BY), share-alike (SA), non-commercial use (NC), and no derivatives (ND). This not only protects the rights of data creators but also provides clear usage guidelines for data users.

\subsection{Growth and Updates of Datasets}
OpenDataLab utilizes automated pipelines and community operations to achieve continuous growth and updates of datasets. Built on a robust big data infrastructure, the automated data collection pipelines facilitate the collection, processing, and uploading of dataset information, ensuring that high-quality and diverse dataset selections are promptly available to users.

Community operations further enhance the platform's development by encouraging community contributions. A mutually beneficial relationship between dataset creators and users establishes an effective closed-loop data supply system. Regular community events, updates on newly released datasets, and active user interactions are key drivers of this process. The latest news can be viewed on the platform's website\footnote{\href{https://opendatalab.com/news}{https://opendatalab.com/news}}.

\section{Data Set Description Language (DSDL)}

\subsection{DSDL Design Goals and Core Architecture}

Datasets are foundational to artificial intelligence research and applications, with their acquisition, dissemination, and utilization rates directly influencing the pace of technological advancement. Throughout the evolution of the AI field, the generation and release of massive datasets have been a key driver of progress. However, these datasets often adhere to diverse definition formats, leading to high costs in their dissemination, integration, and application. To address this issue, we have introduced a new dataset description language: DSDL (Data Set Description Language), with the primary design objectives of universality, portability, and extensibility:

\begin{itemize}
    \item \textbf{Universality}: DSDL provides a unified standard for data representation, applicable across various domains of AI datasets, not limited to specific tasks or fields. This language is designed to articulate datasets of different modalities and structures in a consistent format.

    \item \textbf{Portability}: DSDL ensures high portability of dataset descriptions, allowing description files to be used across different systems and environments without modification, which is crucial for fostering a thriving development ecosystem.

    \item \textbf{Extensibility}: DSDL supports the expansion of dataset descriptions without altering the core standards. Its scope can be extended through libraries or packages while maintaining the long-term stability of the core syntax.

\end{itemize}

\begin{figure*}[h]
    \centering
    \noindent\includegraphics[width=0.7\textwidth]{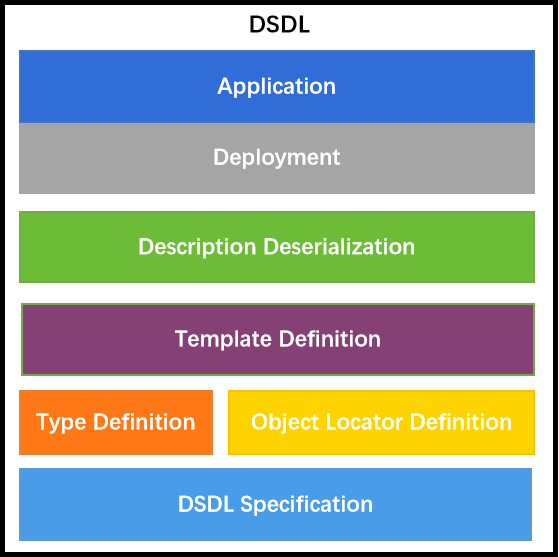}
    \caption{Architecture Diagram of the Data Set Description Language}
    \label{fig:dsdl}
\end{figure*}

As depicted in Figure~\ref{fig:dsdl}, DSDL employs a bottom-up overall architecture. We initially define the DSDL Specification as the foundation of the syntax standard. Building on this, we further delineate DSDL data types and the Object Locator. These building blocks enable researchers to craft dataset descriptions suitable for their AI tasks. The introduction of the Object Locator significantly simplifies the access and utilization of data from multiple sources, whether stored locally, on clusters, or in the cloud. To further facilitate researchers, we have predefined templates for mainstream tasks to streamline dataset usage. A comprehensive syntax parser translates DSDL descriptions into Python objects, completing the functionality of the dataset description language. For more detailed information on DSDL, please refer to the original documentation~\cite{wang2024dsdl}.

\subsection{DSDL User Base}

DSDL serves AI researchers and developers at various levels:

\textbf{Beginners in AI} can quickly understand and utilize mainstream datasets without needing to delve into the complexities of data formats and content, as DSDL provides clear metadata and annotation details.
    
\textbf{Researchers in specific fields} benefit from a unified dataset function interface provided by DSDL, simplifying the use of multiple datasets.

\textbf{Researchers of large models} can efficiently combine and process vast amounts of data across different tasks and modalities, accelerating the training and validation processes of models.

\subsection{DSDL Standardized Datasets}

We have standardized over 100 mainstream datasets, which are readily available for download on the OpenDataLab platform\footnote{\href{https://opendatalab.com/?industry=dsdl\&sort=all}{https://opendatalab.com/?industry=dsdl\&sort=all}}.

Users can easily visualize datasets and explore specific examples, as detailed in the online tutorial.\footnote{\href{https://opendatalab.github.io/dsdl-docs/tutorials/visualization}{https://opendatalab.github.io/dsdl-docs/tutorials/visualization}}. They can also train and deploy models using pre-configured files, significantly simplifying the process and lowering the barriers to entry, as shown in another tutorial.\footnote{\href{https://opendatalab.github.io/dsdl-docs/tutorials/train\_test/openmmlab}{https://opendatalab.github.io/dsdl-docs/tutorials/train\_test/openmmlab}} Additionally, users have the flexibility to combine and utilize standardized datasets as needed.

\section{Open Data Platform and Data Toolkits}
\subsection{Platform Features}

\textbf{Search and Explore}

OpenDataLab provides not only an array of open data resources but also flexible and multi-dimensional dataset retrieval methods, catering to diverse search scenarios. Users can use the global search box for fuzzy searches across various dimensions and filter datasets by data type, label type, task type, and topics. This functionality enables users to identify similar datasets, fostering a more comprehensive collection. The platform also facilitates the discovery of the most recent datasets by offering sorting options based on popularity and update time. Additionally, a random recommendation function is available for broader exploration scenarios.

\textbf{Preview and Describe}

OpenDataLab employs Data Cards, comprising a README and a Metafile, and a standardized DSDL for dataset definition. These components allow users to grasp the dataset's essentials, facilitating informed decisions regarding its applicability.

OpenDataLab offers an extensive array of data visualization techniques to support visual tasks such as image classification, segmentation, object detection, tracking, and OCR recognition. Additionally, it facilitates statistical and analytical visualization of distribution metrics, including file format, size, and resolution. This comprehensive approach provides users with an intuitive understanding of dataset attributes.

\begin{figure*}[h]
    \centering
    \noindent\includegraphics[width=1.0\textwidth]{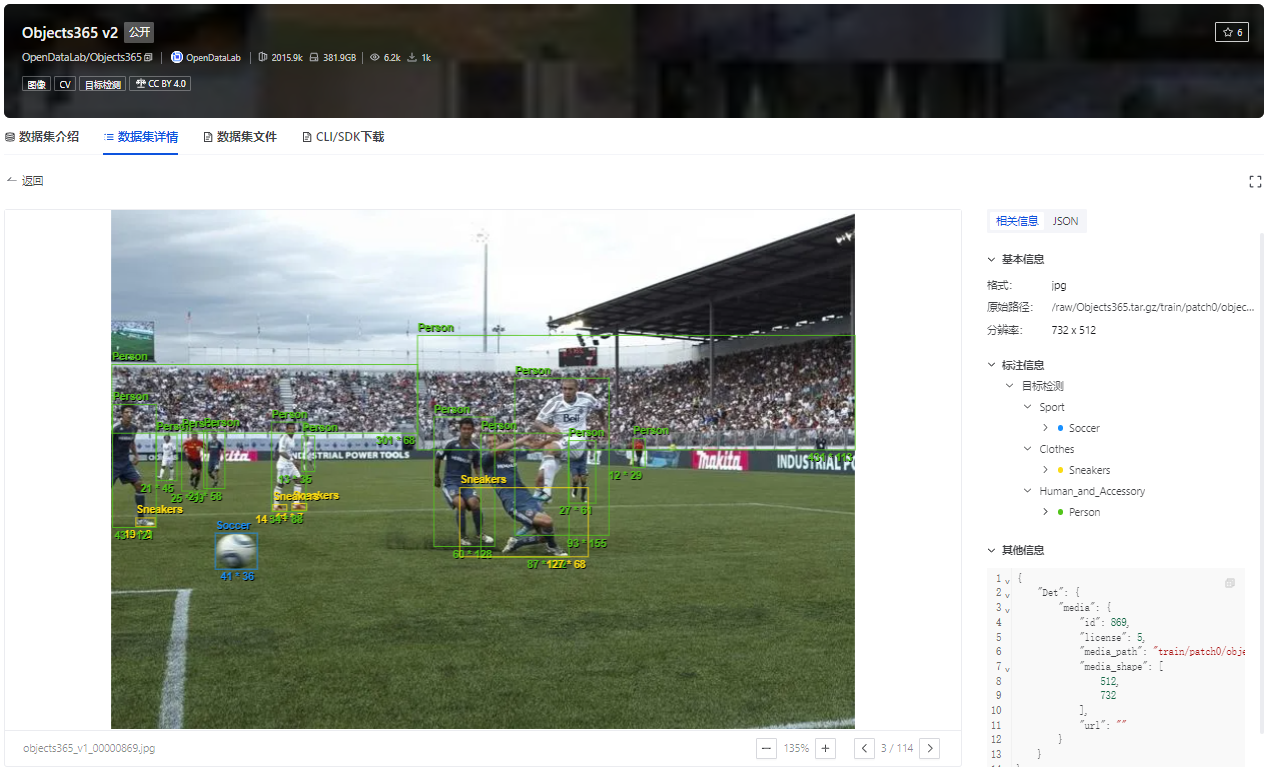}
    \caption{OpenDataLab Preview}
    \label{fig:figure6}
\end{figure*}

The following shows a visualization example~\ref{fig:figure6}.

\textbf{Host and Download}

OpenDataLab offers free dataset hosting with multiple download options. Users can upload dataset files via a web browser or use the platform's Command Line Interface (CLI) or Python SDK. Datasets can be downloaded using web browser or code tools by network. Users are encouraged to identify the required dataset and verify its suitability through meta-information, introductory details, and visual examples which prevents the time-consuming download of large-scale data.

\subsection{OpenDataLab CLI and SDK}
OpenDataLab CLI and SDK are provided to platform users as sub-functions of the OpenXLab [1] dataset module. OpenDataLab CLI and SDK provide users with different usage methods, including search, query, view details, create data sets and data download functions, and support users to contribute data sets, search and download data resources.

\textbf{Command Line Interface for Datasets}

The OpenDataLab CLI offers command line tools for dataset publishers and users, with the command line structure specified as follows:

\begin{tcolorbox}
\begin{verbatim}
openxlab dataset <subcommand >  [options and parameters]
\end{verbatim}
\end{tcolorbox}

\begin{itemize}
\item <openxlab dataset>: refers to the name of the CLI tool of OpenDataLab.

\item <subcommand>: Specify the additional sub command to perform the operation, such as get or download.

\item <options and parameters>: Specify options or API parameter options used to control CLI behavior. The option values can be numbers, strings, JSON structure, etc.
\end{itemize}

The OpenDataLab CLI command line tool includes several functions that support users in hosting, listing, downloading, and managing datasets on OpenDataLab.

\textbf{Python SDK for Datasets}

The OpenDataLab Python SDK offers users a programmatic approach to managing and operating datasets, enhancing flexibility across various system environments and data usage contexts. The features provided by the OpenDataLab Python SDK are nearly identical to those of the CLI, thereby simplifying the management and utilization of datasets.

\subsection{Data Labeling Tools}

\textbf{LabelU - Labeling tools for ML }

LabelU is a data labeling tool library specially designed for the field of artificial intelligence and machine learning. It encapsulates over 15 labeling tools for image, audio, and video tasks. 

LabelU's image tools enable the labeling of boxes, points, lines, polygons, and 3D boxes. Its video and audio tools provide support for segmentation, classification, and transcription, thereby facilitating the precise processing of complex visual and auditory information.

LabelU offers customizable tool configurations, allowing a flexible combination of multiple tools to suit diverse user needs. With its open-source code and local installation support\footnote{\href{https://github.com/opendatalab/labelU}{https://github.com/opendatalab/labelU}}, LabelU fosters convenient, precise, and efficient data labeling.

\begin{figure*}[ht]
    \centering
    \noindent\includegraphics[width=1.0\textwidth]{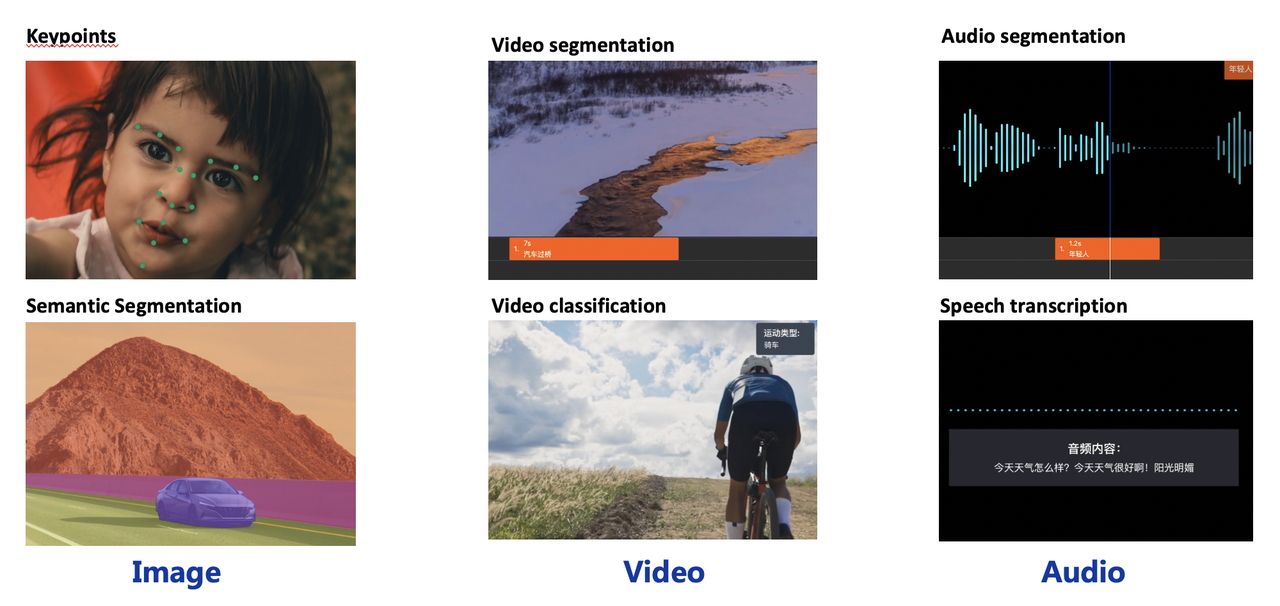}
    \caption{Labeling Case by LabelU}
    \label{fig:figure4}
\end{figure*}

The cases annotated using LabelU are illustrated in Figure~\ref{fig:figure4}.

\textbf{LabelLLM - Labeling tools for AGI}

LabelLLM is an intelligent dialogue labeling tool designed to provide high-quality annotated data for large models. It offers capabilities for labeling and reviewing conversation data, thereby enabling precise processing of NLP and multi-modal data. The platform includes tools for classifying and annotating entire conversations or individual questions, as well as for sequencing each response within a conversation.

LabelLLM allows for flexible tool combinations and task parameter configurations. It is applicable in a variety of contexts, such as question-answer collection, preference gathering, and dialogue evaluation necessary for large models. It provides an efficient toolkits for data preparation in the fine-tuning and RLHF stages of large model development.

\begin{figure*}[h]
    \centering
    \noindent\includegraphics[width=1.0\textwidth]{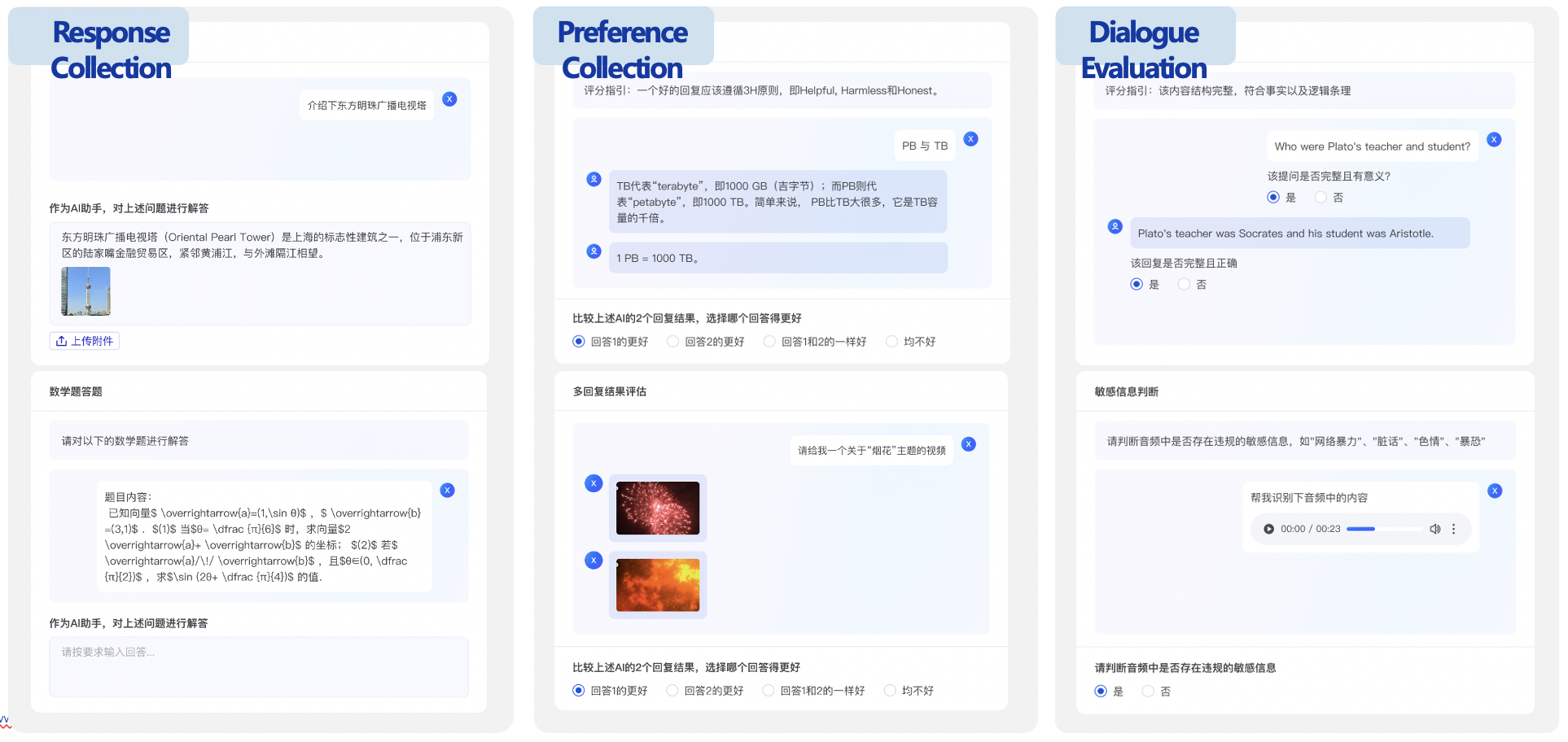}
    \caption{Labeling Case by LabelLLM}
    \label{fig:figure5}
\end{figure*}

Figure~\ref{fig:figure5} illustrates examples of data annotated using LabelLLM.

\section{Use Case}
\begin{figure*}[h]
    \centering
    \noindent\includegraphics[width=1.0\textwidth]{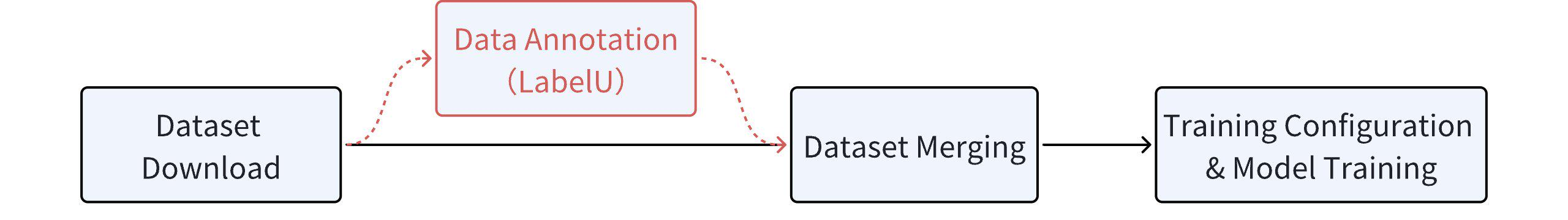}
    \caption{Use case by using OpenDataLab}
    \label{fig:figure7}
\end{figure*}

Using the object detection task as an illustrative example, we demonstrate the convenience offered by OpenDataLab throughout the entire research and development process, as depicted in the Figure~\ref{fig:figure7}. 

\subsection{Dataset Download}

To train an object detection model, the initial step involves acquiring the relevant data. OpenDataLab streamlines this process, enabling users to effortlessly query, select, and download the desired dataset. 

\begin{tcolorbox}
\begin{verbatim}
$ openxlab dataset get --dataset-repo OpenDataLab/PASCAL_VOC2007
$ openxlab dataset get --dataset-repo OpenDataLab/COCO_2017
\end{verbatim}
\end{tcolorbox}

\subsection{Dataset Merge and Model Training}

The PASCAL VOC2007 and COCO2017 datasets can be downloaded using above command. However, their original data formats are not consistent, can not direct merging. We shall use DSDL to standardized the datasets for combined training. Our support extends to Pytorch training and the MMDetection framework by OpenMMLab\footnote{\href{https://github.com/open-mmlab}{https://github.com/open-mmlab}}, with detailed usage instructions to follow.

\textbf{Pytorch Implementation (Merge by SDK)}

For Pytorch users, using the merged data set to train the model can directly merge the data through the SDK. Eliminate a lot of handwritten code alignment and merging process.

\begin{tcolorbox}
\begin{verbatim}
from dsdl.dataset import DSDLDataset, DSDLConcatDataset

voc_yaml_path = f"path_to_your_voc_data/dsdl/set-train/train.yaml"
ds_voc = DSDLDataset(dsdl_yaml=voc_yaml_path)

coco_yaml_path = f"path_to_your_coco_data/dsdl/set-train/train.yaml"
ds_coco = DSDLDataset(dsdl_yaml=coco_yaml_path)

train_data = DSDLConcatDataset([ds_voc, ds_coco])

# Your Traning Code ...
\end{verbatim}
\end{tcolorbox}
\vspace{15pt}

\textbf{OpenMMLab (Merge by config)}

For OpenMMLab users, data merging can be achieved directly through ConcatDataset in the configuration file without any changes to the training code.

\begin{tcolorbox}
\begin{verbatim}
python tools/train.py config/dsdl/voc2007_coco2017.py

#following is the content of voc2007_coco2017.py
 dataset=dict(
    type="ConcatDataset",
    datasets=[
        dict(
            type="DSDLDetDataset",
            data_root="path to your data",
            ann_file="dsdl/set-train/train.yaml",
            ...
        ),
        dict(
            type="DSDLDetDataset",
            data_root="path to your data",
            ann_file="dsdl/set-train/train.yaml",
            ...
        ),
        ...
)

\end{verbatim}
\end{tcolorbox}

\subsection{Visualization and Labeling}

LabelU equips users with tools for visual data analysis and label modification, fostering clear visualization of raw images and detailed annotations, and enabling customization of labeling tasks based on user requirements. 

Consequently, OpenDataLab provides an integrated solution for data download, training, and modification, thereby augmenting the efficiency and effectiveness of AI research.

\section{Related Work}
\subsection{Unified Data Standards}
Deep Lake\cite{DeepLake} is a data lake optimized for deep learning, capable of interfacing with various storage systems and enhancing data operations with its advanced query language. It streamlines data handling for machine learning with features like version control, efficient data streaming, and a visualizer for large datasets. 

Hugging Face\cite{Huggingface} is a hub for accessing and sharing a wide array of datasets tailored for natural language processing tasks. It simplifies the process of finding, using, and contributing to datasets, offering tools for easy dataset manipulation and collaboration within the NLP community. While platforms like Hugging Face Data provide an extensive collection of datasets for natural language processing tasks, they do not address the unification of AI data formats. A standardized specification that enables interoperability among various tasks and even different modalities is crucial for advancing multimodal research.

Data Set Description Language (DSDL) is designed precisely for this purpose. DSDL not only standardizes diverse data types through a unified specification but also boasts robust language extensibility, progressively encompassing a broader spectrum of AI data types. This approach lays a solid foundation for multimodal AI research by facilitating interoperability and interconnectivity of data.

\subsection{Data Platform}
\textbf{Data Tools}
The advancement of artificial intelligence has prompted an influx of open-source labeling tools, exhibiting a broad spectrum but often constrained by their application range, tool functionality, and interoperability. These constraints impede their adaptability to the swiftly progressing AI sector and its diverse needs. For instance, CVAT\cite{CVAT}, an Opencv's web-based image labeling tool, excels in various image annotations but lacks in text and audio data support. Likewise, Labelme\cite{Labelme}, a graphical interface CV annotation software from MIT, primarily supports single-task annotations, lacking multi-task integration.

The onset of the large model era has amplified the demand for fine-tuning and RLHF, necessitating tools capable of handling dialogue labeling and configuration scenarios. The existing tool, Open Assistant\cite{Open-Assistant}, is an open-source large model labeling tool. The functions of Open Assistant include marking prompts, adding reply messages, and editing assistant replies. However, its functionality is confined to text types, and it lacks in multi-modal data comprehension and task configuration capabilities, rendering it ill-equipped for the diverse multi-modal scenarios demanded by the large model industry.

\textbf{Command Line Interface for Datasets}

The Well-known open data platforms at home and abroad were selected to compare OpenDataLab, Kaggle\cite{Kaggle}, Huggingface, Paperwithcode\cite{Paperswithcode}, ModelScop\cite{Modelscope}, and BAAI\cite{BAAI} from three aspects: platform functions, platform content, and platform standardization.

Established in 2010, Kaggle is a globally recognized data science competition platform offering numerous public datasets and fostering community discussions. However, its scope of visual, audio data, and large model-related data remains limited. Hugging Face, introduced in 2016 amidst the rise of deep learning, is an open platform specializing in natural language processing. It houses a plethora of open-source models and diverse linguistic data, but lacks in other modalities besides text.

The growth of artificial intelligence has sparked the emergence of open AI platforms like ModelScope and BAAI. ModelScope, a product of Alibaba Damo Academy, operates as a model-as-a-service platform, offering developers comprehensive model services and large model data hosting. BAAI, under the Beijing Zhiyuan Artificial Intelligence Research Institute, serves as a data platform. However, both platforms are limited in data scale and richness of data information details.

OpenDataLab, established in 2022, is an open AI data platform providing a superior data exploration and download experience. It features comprehensive search, filtering, and online visualization of media data annotation information, complemented by a user-friendly GUI design.

The comparative details are shown in Table~\ref{tab:tab1}:

\begin{table}[htbp]
    \centering
    \caption{Comparison of Data Platforms.}
    \label{tab:tab1}
    \tiny
    \begin{tabularx}{\textwidth}{c|l|c|c|c|c|c|c}
        \toprule
        \multicolumn{2}{c|}{\textbf{Dimension}} & \textbf{OpenDataLab} & \textbf{\makecell{Kaggle\\ Datasets}} & \textbf{\makecell{Hugging Face\\ Datasets}} & \textbf{Paperwithcode} & \textbf{ModelScope} & \textbf{BAAI} \\
        \midrule
        \multirow{5}{*}{{\textbf{\makecell{Platform\\ Features}}}} & \textbf{Powerful data filtering} & \Checkmark & \Checkmark & \Checkmark & \Checkmark & $\bigtriangleup$ & $\bigtriangleup$ \\
        & \textbf{Various data visualization} & \Checkmark & \Checkmark & \Checkmark & \XSolidBrush & $\bigtriangleup$ & \XSolidBrush \\
        & \textbf{Host and fast download} & \Checkmark & $\bigtriangleup$ & $\bigtriangleup$ & \XSolidBrush & \Checkmark & $\bigtriangleup$ \\
        & \textbf{Strong management ability} & $\bigtriangleup$ & \Checkmark & \Checkmark & \XSolidBrush & $\bigtriangleup$ & $\bigtriangleup$ \\
        & \textbf{High consistency across clients} & \Checkmark & \Checkmark & $\bigtriangleup$ & \Checkmark & \Checkmark & $\bigtriangleup$ \\
        \midrule
        \multirow{4}{*}{{\textbf{\makecell{Platform\\ Datasets}}}} & \textbf{Large scale open datasets} & \Checkmark & \Checkmark & \Checkmark & \Checkmark & \XSolidBrush & \XSolidBrush \\
        & \textbf{Wide applicability} & \Checkmark & $\bigtriangleup$ & $\bigtriangleup$ & $\bigtriangleup$ & $\bigtriangleup$ & \XSolidBrush \\
        & \textbf{High quality data} & \Checkmark & \Checkmark & $\bigtriangleup$ & \XSolidBrush & $\bigtriangleup$ & $\bigtriangleup$ \\
        & \textbf{Timely updated} & \Checkmark & $\bigtriangleup$ & \Checkmark & $\bigtriangleup$ & \XSolidBrush & \XSolidBrush \\
        \midrule        
        \multirow{3}{*}{{\textbf{\makecell{Platform\\ Standardization}}}} & \textbf{Standardization in meta} & \Checkmark & \Checkmark & \Checkmark & $\bigtriangleup$ & \Checkmark & $\bigtriangleup$ \\
        & \textbf{Standardization in Labels} & \Checkmark & \XSolidBrush & $\bigtriangleup$ & \XSolidBrush & $\bigtriangleup$ & \XSolidBrush \\
        & \textbf{Flexibility in interface} & \Checkmark & $\bigtriangleup$ & \Checkmark & \Checkmark & \Checkmark & $\bigtriangleup$ \\
        \bottomrule
        \multicolumn{8}{l}{\Checkmark: Strong features. \XSolidBrush: Few features or Not supported. $\bigtriangleup$: General features. }\\
    \end{tabularx}
\end{table}

\section{Conclusion}
OpenDataLab, an open data platform, is committed to providing high-quality datasets for artificial intelligence, fostering data sharing, collaboration, and comprehensive support for global AI initiatives. It supplies a broad spectrum of training data, pertinent to a range of models including language, multi-modal, video, three-dimensional, and speech. The data finds extensive applications across various industries such as autonomous driving, smart healthcare, and digital cities.

OpenDataLab has introduced a Data Set Description Language (DSDL), enabling a unified representation of datasets across various formats and modalities. OpenDataLab also develops intelligent labeling tools and extensive multi-modal datasets to facilitate efficient cross-modal and cross-task data acquisition, labeling, and modification, thus enhancing data preparation efficiency by 30\%. 

OpenDataLab aims to pioneer a new paradigm for data preparation in the AI industry, thereby catalyzing the progression of a new era of large-scale models.

\newpage

\bibliographystyle{plain}
\bibliography{opendatalab.bib}

\end{document}